\documentclass[two column,pra,showpacs]{revtex4}

\usepackage{graphicx}
\usepackage{dcolumn}
\usepackage{bm}
\usepackage[latin1]{inputenc}
\usepackage{hyperref}

\begin{document}

\preprint{}

\title{Three-vortex configurations in trapped Bose-Einstein condensates}

\author{J.~A.~Seman$^1$}\email{jorge@ursa.ifsc.usp.br}
\author{E.~A.~L.~Henn$^1$}
\author{M.~Haque$^2$}
\author{R.~F.~Shiozaki$^1$}
\author{E.~R.~F.~Ramos$^1$}
\author{M.~Caracanhas$^1$}
\author{P.~Castilho$^1$}
\author{C.~Castelo Branco$^1$}
\author{P.~E.~S.~Tavares$^1$}
\author{F.~J.~Poveda C.$^1$}
\author{G.~Roati$^3$}
\author{K.~M.~F.~Magalh\~{a}es$^1$}
\author{V.~S.~Bagnato$^1$}
\affiliation{$^1$Instituto de F\'{\i}sica de S\~{a}o Carlos -- USP. C.P. 369, S\~{a}o Carlos -- SP - Brazil -- 13560-970\\ $^2$ Max Planck Institute for Physics of Complex Systems N\"othnitzer Strasse 38, 01187 Dresden, Germany \\ $^3$LENS and Dipartimento di Fisica, Universita di Firenze, and INFM-CNR, Via Nello Carrara 1, 50019 Sesto Fiorentino, Italy}

\begin{abstract}

  We report on the creation of three-vortex clusters in a $^{87}Rb$ Bose-Einstein condensate by oscillatory excitation of the condensate.  This procedure can create vortices of both circulation, so that we are able to create several types of vortex clusters using the same mechanism.  The three-vortex configurations are dominated by two types, namely, an equilateral-triangle arrangement and a linear arrangement.  We interpret these most stable configurations respectively as three vortices with the same circulation, and as a vortex-antivortex-vortex cluster.  The linear configurations are very likely the first experimental signatures of predicted stationary vortex clusters.

\end{abstract}

\maketitle

\section{Introduction}

Quantized vorticity being a key feature of
superfluidity, vortex dynamics and configurations have long been central
topics in the study of Bose-Einstein condensates (BECs).  In the context of
trapped atomic BECs, vortex dynamics has enjoyed a resurgence of interest
because of new physical effects associated with the trap geometry.

Vortices have been produced in BECs by transferring angular momentum either by
manipulation of the quantum phase and the hyperfine state of the atoms
\cite{firstvortex} or directly by mechanical means \cite{vortexformation,
PRA}.  The majority of subsequent experiments have focused on externally
rotated condensates, where vortices of the same circulation sign arrange into
vortex lattices \cite{vortexcluster}.  On the other hand, in BECs without
external rotation, one can have vortices of both circulation signs coexisting.
In a trap, such combinations display fundamentally new effects, such as
stationary structures found in theoretical studies \cite{crasovan, mottonen,
  pietila}.  For example, in pancake-shaped BECs, a stable linear
configuration exists for vortex tripoles if the interaction is large enough
\cite{mottonen,pietila}.  This linear arrangement contains an anti-vortex at
the trap center and two vortices equally spaced on either side of the
anti-vortex.

The purpose of this paper is to report experimental evidence for the stability
of specific three-vortex configurations in a trapped BEC.  We create
three-vortex states through an external oscillatory perturbation in a
$^{87}Rb$ BEC held in a cigar-shaped magnetic trap.  Our analysis of the
three-vortex configurations indicate the existence of stable linear
structures.  We interpret these as the vortex tripole structures of
Refs.\ \cite{mottonen, pietila} described above.  We also find structures
where the three vortices form an approximately equilateral triangle.  We
interpret these as three vortices with the same circulation.

In the absence of extensive experimental results on few-vortex dynamics and
configurations in non-rotating trapped BECs, our results open up an intriguing
direction of vortex studies.  By realizing the linear configuration in a
cylindrical geometry, we demonstrate that the phenomenon of stationary vortex
cluster configurations is robust well beyond the two-dimensional (2D) or
pancake limit.  In addition, the study of the dynamics and coexistence of
vortices and anti-vortices is an essential ingredient for understanding
evolution to turbulence as well as decay processes.  More generally, interest
in the interplay between vortex dynamics and confinement effects extends
beyond cold-atom physics, since analogous issues with magnetic
vortex/anti-vortex effects in ferromagnetic microstructures have attracted a
great deal of attention in recent years \cite{magneticvortices}.

\section{Experimental results}

The experimental setup and the vortex
excitation procedure are described in detail elsewhere \cite{BJP,
PRA}. Briefly, we use a $^{87}Rb$ BEC with $1\times10^5$ atoms held in a cigar
shaped magnetic trap with frequencies $\omega_z=2\pi\times23$ and
$\omega_r=2\pi\times207$ Hz. Vortices are generated by applying on the BEC an
oscillating quadrupolar magnetic field whose amplitude is much smaller than
the trap fields. As a function of field amplitude we observe an increasing
number of vortices formed, up to the limit where a vortex tangle is observed
\cite{turbulence}.

For this study, we focus on images where a clear 3-vortex configuration is
observed within the condensate core. Therefore, we have restricted the
amplitude and frequency of the excitation for generating only this kind of
structure.  The excitation time was fixed at 20~ms, followed by an
equilibrating time of 20~ms before releasing the atoms for free expansion.
This is important because different excitation times can lead to different
physical situations. Vortices are produced while the BEC is still held in the
trap.  For our analysis we will assume that the in-trap structure is
maintained during the 15~ms of time-of-flight. In fact, this is a widely used
assumption in experimental vortex studies.
Of course, it is possible that in some of the images two of the density dips
might be the two ends of a single vortex line, rather than two distinct
vortices.  This situation may introduce ``false counts" in our statistics and
is a limitation of our procedure for detecting and counting vortices.

\begin{figure}
\centering
 \includegraphics[scale=0.35]{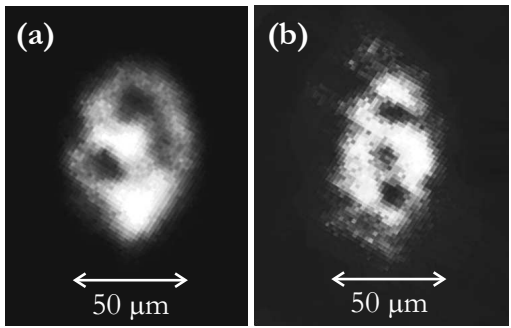}
 \caption{Atomic optical density images showing configurations of stable vortices forming (a) an equilateral triangle, or (b) a linear array. Images were taken after 15 ms of free expansion.}
\label{fig:three}
\end{figure}

\begin{figure}
\centering
 \includegraphics[scale=0.4]{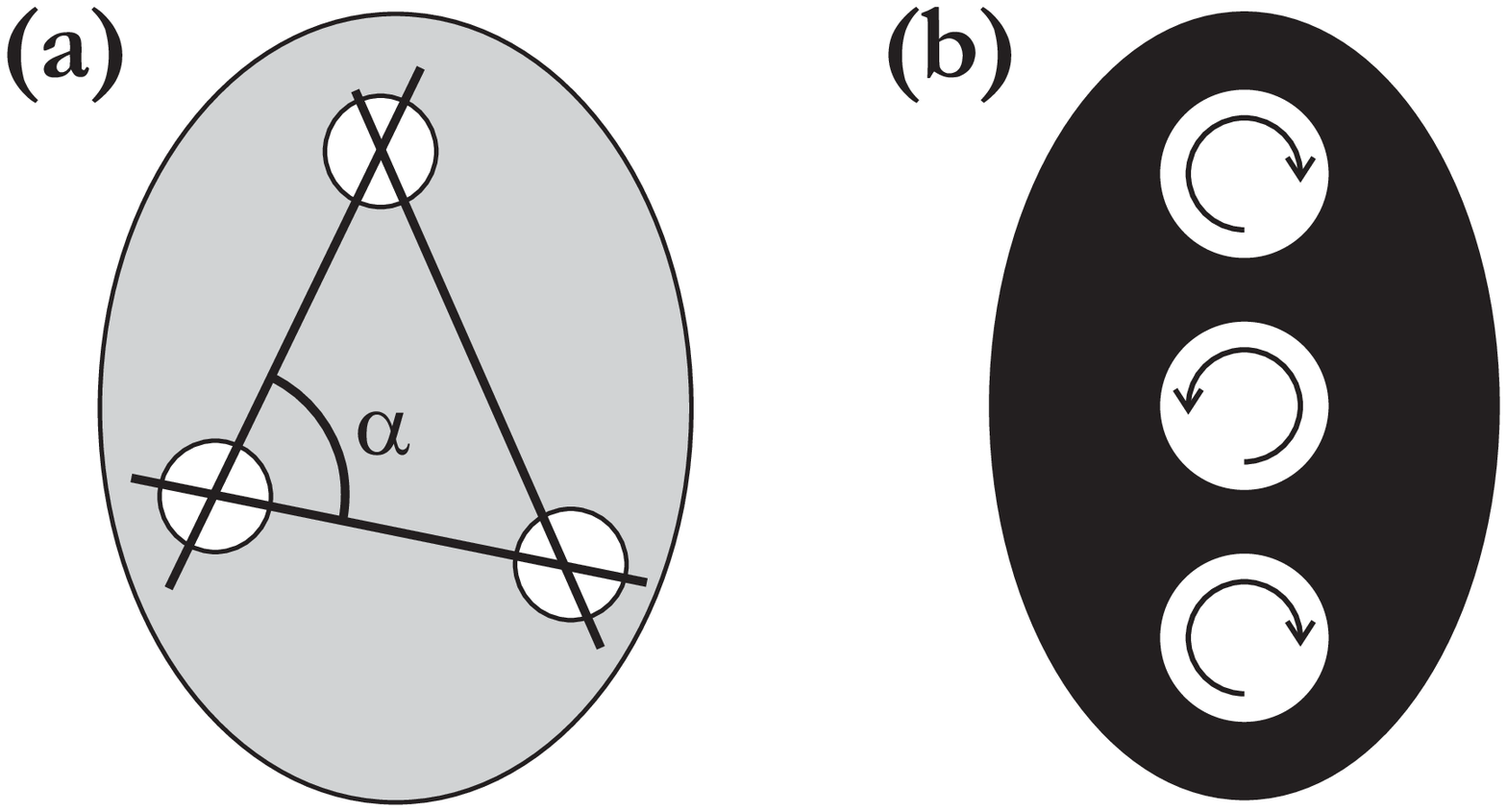} 
 \caption{(a) Definition of the largest internal angle $\alpha$.
(b) Schematics of the tripole configuration of vortices, arrows indicate the vortex circulation direction.}
\label{fig:schematics}
\end{figure}

We observe predominantly two types of configurations. The vortices are either
distributed as a near-equilateral triangle, as in Fig.~\ref{fig:three}(a), or
as a near-linear array, as in Fig.~\ref{fig:three}(b).  We analyze the
geometrical distribution by recording the largest internal angle, $\alpha$, of
the triangle whose vertices are the vortex positions, as sketched in
Fig.~\ref{fig:schematics}(a). The histogram of Fig.~\ref{fig:angle} summarizes
our results of more than 60 independent measurements, showing the relative
frequencies of observed values of $\alpha$. Two types of stable configurations
have a clear prominence over the others, namely, an equilateral triangle type
where the angle $\alpha$ is around $60^o$, and a linear array where
$\alpha\sim180^o$. 

It is important to point out that, for the present conditions of the excitation, 
vortices are mainly nucleated along the radial direction of the condensate, 
perpendicular to the long axis of the cloud. For the case of the linear array, the 
orientation of the cluster is preferentially aligned with the long axis of the 
condensate.

\section{Interpretation}

We will interpret our data using considerations of a quasi-2D (pancake-shaped)
geometry.  While this is a crude approximation to our condensate geometry,
going beyond this approach is difficult because results on few-vortex dynamics
in our geometry are rare.

\begin{figure}
\centering
 \includegraphics[scale=0.3]{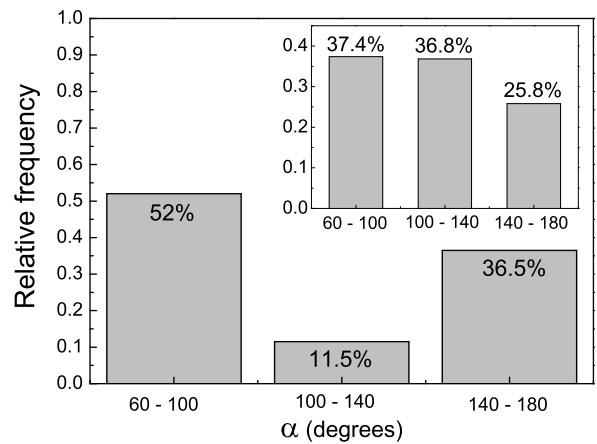}
 \caption{Observed relative frequency of 3-vortex configurations as a function
 of the angle $\alpha$. The inset shows the expected distribution of $\alpha$
 when the vortices are distributed at random positions.}
\label{fig:angle}
\end{figure}

\begin{figure}
\centering
 \includegraphics[width=0.95\columnwidth]{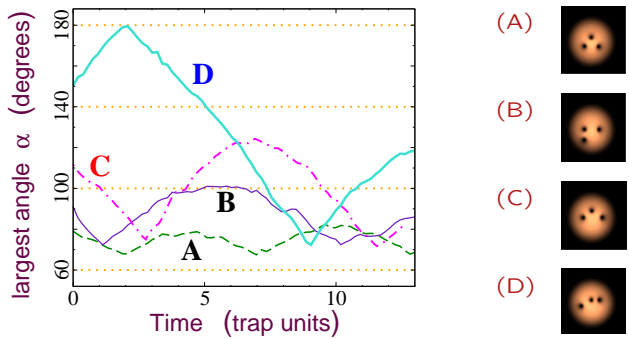}
 \caption{(Color online.)  Evolution of the largest angle $\alpha$, in
   Gross-Pitaevskii simulations starting from various three-vortex
   configurations in a circularly trapped two-dimensional BEC. Initial
   configurations are shown on right.   
} 
\label{fig:simulations}
\end{figure}

In the inset of Figure \ref{fig:angle}, we show the
expected distribution of the largest angle $\alpha$ when the three vortices
are distributed at random positions within the magnetically trapped BEC core, 
taken to be an ellipse with aspect ratio $\sim1.5$.  The intermediate
$\alpha\in[100^o,140^o]$ arrangement would be expected to be almost as
numerous as the equilateral-like $\alpha\in[60^o,100^o]$ arrangement.  To
explain the fact that we get much fewer intermediate configurations, we
consider the \emph{dynamics} of three vortices with the same circulation.  For
simplicity, our discussion and the simulations of Figure \ref{fig:simulations}
use a circular trap, but this is sufficient for the present discussion.

Three vortices placed symmetrically around the trap center ($\alpha=60^o$) in
a nonrotating condensate simply precess around the center, maintaining the
equilateral configuration.  If the initial configuration is moderately
distorted from this symmetric configuration, the dynamics of the triangle
shape is also moderate, so that $\alpha$ does not vary too much.  Thus, a
significant fraction of the configurations starting with
$\alpha\in[60^o,100^o]$ stays within this range during the precession
dynamics (Fig.\ \ref{fig:simulations}A,B).

However, when the initially created vortex positions are such that
$\alpha\in[100^o,140^o]$, the triangle formed by their positions changes shape
as the vortices precess, so that the oscillations of $\alpha$ takes it in and
out of the $\alpha\in[60^o,100^o]$ range (\emph{e.g.},
Fig.\ \ref{fig:simulations}C).  Thus the preponderance of near-equilateral
configurations in our experiment reflects the fact that there are many
$\alpha\in[60^o,100^o]$ configurations which are `stable' in the sense of
maintaining the triangle shape while precessing.  Clearly, the experimental
data should be thought of as corresponding to random times in
Fig.\ \ref{fig:simulations}.

While the equilateral arrangement is reminiscent of the Abrikosov lattice
observed by many authors \cite{abrikosov}, and of the three-vortex case
studied in Refs.\ \cite{aftalion} and \cite{garcia&garcia}, our measurements
explore the rather different physics of non-rotating condensates.  In an
Abrikosov lattice, the equilateral triangle does not distort and stays
stationary in the cororating frame, and the vortices do not undergo mutual
rotations and independent precession-dynamics, as in the present case.  Our
results reveal that the equilateral arrangement also plays a special role in
the situation without external rotation, through maintaining its shape during
precession.

From similar arguments (Fig.\ \ref{fig:simulations}D), if all vortices have
the same circulation direction, we would expect a smaller frequency of
near-linear configurations with $\alpha\in[140^o,180^o]$, compared to the
random-distribution case of Figure \ref{fig:angle} inset.  In our experimental
data, however, the linear array is more frequent than the random-distribution
prediction.  This indicates that in the linear configurations the vortex
circulations are not all of the same sign.

Refs.~\cite{mottonen,pietila} have shown that a tripole configuration like the
one shown in Fig.~\ref{fig:schematics}(b) can be a stable configuration.  For
a pancake-shaped condensate, the vortex tripole is metastable when the
interaction strength is above a certain value \cite{pietila}.  In this
situation, the interaction strength is expressed by the effective 2D coupling
parameter $\tilde{g}$, defined as

\begin{equation}\label{eq:g}
\tilde{g}=\sqrt{8\pi}\frac{Na}{a_{\perp}},
\end{equation}
where $a$ is the scattering length, $N$ is the number of particles, and
$a_{\perp}$ is the oscillator length in the direction perpendicular to the
plane considered to be quasi-2D.  Following Ref.~\cite{pietila} the condition
for the tripole configuration to be stable is $\tilde{g}\geq108$.  In our
experimental geometry, the perpendicular direction is one of the radial
directions, $a_{\perp}=a_r=\sqrt{\hbar/m\omega_r}$, so that
$\tilde{g}\approx200$.
Thus, from quasi-2D arguments, we expect the linear tripole to be metastable
in our setup, which supports our interpretation that the predominance of
linear configurations is due to the tripole of Refs.~\cite{mottonen,pietila}.
The fact that our setup is not really quasi-2D indicates that the results of
Ref.~\cite{pietila} are valid well beyond the pancake-shape limit.

It is expected \cite{pietila} that the tripole survives for a time of the
order of $\omega_0^{-1}$, which for our case is about 10~ms, well within the
20~ms of waiting time before releasing the atoms from the magnetic potential.
Thus tripole decay has a chance to reduce the number of observations of
tripole configurations, as discussed further below. In fact, 20~ms of waiting 
time seems to be the best time to clearly observe the vortices within the 
cloud and, for this reason, it has been chosen in our experiment. For a very 
long waiting time after excitation, observation of collinear 3-vortex 
configurations seems to be scarcer but fluctuations do not allow us to be more 
quantitative in this regard. 

The experimental method used here to produce vortex states has basically the
same probability to produce both circulation sign vortices. Such equal chances
come from the fact that the oscillatory field introduced by the extra coils
modulates the equilibrium position as well as the overall trap shape
introducing features that result in a vortex state of both circulation signs,
as discussed in Ref.~\cite{PRA}. The equal chances for vortex and anti-vortex
formation implies that in many of the observed distributions a collection of
two vortices and one anti-vortex can occur.  Since such a combination has a
metastable linear structure, this would explain the prominence of near-linear
structures in Figure \ref{fig:angle}.

\section{Relative abundance of configurations}

We now provide back-of-the-envelope estimates for the relative abundance of
the near-equilateral and near-linear configurations.  
If vortex and anti-vortex production are equally likely, we would expect that
the probability $P_{line}$ to excite a mixture containing one vortex with
circulation opposite to the other two would be higher than the probability
$P_{triangle}$ to produce three vortices with the same sign.  Considering the
data presented in Fig.~\ref{fig:angle}, we have $P_{line}\approx0.37$ and
$P_{triangle}\approx0.52$.  Assuming the production of a sequence of vortices
to be independent events, we would have the probability of having three
vortices of the same sign as 0.25 while for having one of opposite sign among
them as 0.75. That would produce a ratio $1:3$ and not $1:0.7$ as observed in
our experiment. This mismatch is probably due to the complex dynamics
occurring after vortex nucleation.  Vortex tripoles have more decay
mechanisms, such as a vortex/anti-vortex pair annihilation or through one
vortex migrating to the borders of the condensate generating surface modes,
and leaving behind a vortex dipole \cite{pietila}.  The latter process is
known to happen at time scales $\sim\omega_0^{-1}$ \cite{pietila}.
Considering an exponential decay for the tripole configurations, with time
constant $\omega_0^{-1}\approx10$ ms, we get for our waiting time of 20 ms a
reduction by a factor of $e^{-2}\approx0.135$.
Taking all this into account, if a configuration with three vortices is
initially produced, the probability of observing the triangle remains as 0.25,
the probability of observing a tripole is 0.10 and the probability of
observing a dipole is 0.65.  Considering only three-vortex observations, we
have $P_{triangle}\approx0.715$ and $P_{line}\approx0.285$.  This is much
closer to the observed values in our experiment (Fig.~\ref{fig:angle}).
This suggests that the data in Fig.~\ref{fig:angle} reflects greater stability
of the same-circulation situation compared to the vortex/anti-vortex/vortex
configuration.

\section{Discussion}

In summary, we have observed 3-vortex cluster
configurations where two different types are the most frequent: the
equilateral triangle geometry and the linear array.  We have argued that the
triangular cluster type is formed predomniantly by vortices with the same sign
of circulation while the linear array is numerous because of configurations
where the middle vortex has opposite sign.

The present experiments are loosely related to the work of Ref.~\cite{merging}
where different vortex configurations were created by merging multiple BECs.
Almost linear or triangle structures can also be seen in their data.  Our
results complement those of Ref.~\cite{merging}, using a quite different
vortex formation mechanism.  More recently, the same group has produced vortex
dipoles in a non-rotating superfluid by forcing a quasi-2D BEC to flow around
a repulsive Gaussian obstacle \cite{neely}.

Vortex-antivortex dynamics in confined geometries is of broader interest than
superfluid gases, since similar dynamical issues are currently under intense
study in the magnetic microstructure context \cite{magneticvortices}.  Within
the field of cold atomic gases, our experiment features a distinct manner for
creating vortices of both circulation in the same condensate.  Our results
highlight several poorly understood aspects of vortex physics, in particular
involving non-equilibrium dynamics.  Since the imaging is performed after the
released condensates have expanded and have undergone aspect ratio inversion,
there are several issues related to the expansion dynamics. It is not known in
any detail whether or not the presence of vortex clusters affects the
inversion dynamics, or how the vortex configurations are rearranged as the
cloud undergoes anisotropic expansion.

Another issue raised by the current experiments is the stability of the
tripole configurations in a cylindrical rather than pancake-shaped condensate,
since the theoretical studies \cite{crasovan, mottonen, pietila} concern flat
condensates.  The abundance of the linear tripole configuration seen here
suggests that the scissor oscillations used to excite the condensate may
stabilize the tripole configuration, perhaps counteracting the dynamical
instability discussed in Ref.~\cite{pietila}. These effects are clearly of
basic interest but are yet to be theoretically analyzed in detail.  In
addition, investigation of the dynamics of superfluids containing both
vortices and anti-vortices may be a fruitful alternative way to understand
basic decay mechanisms of quantum turbulence and other poorly understood
phenomena in atomic BECs.

Special thanks to C. Salomon and W. D. Phillips for fruitful discussion. This work was supported by FAPESP (research program CEPID/CEPOF), CAPES and CNPq (research program INCT).

\end{document}